\documentclass{article}%
\usepackage{amssymb}
\usepackage{amsmath}%
\usepackage{amsfonts}%
\usepackage{graphicx}
\providecommand{\U}[1]{\protect\rule{.1in}{.1in}}
\begin{document}

\title{Subdynamics of fluctuations in an equilibrium classical many-particle\\system and generalized linear Boltzmann and Landau equations}
\author{Victor F. Los
\and Institute for Magnetism, Nat. Acad. Sci. and Min. Edu. Sci. of Ukraine
\and Vernadsky Blvd. 36-b, 03142, Kiev, Ukraine
\and E-mail address: victorflos43@gmail.com}

\begin{abstract}
New exact completely closed homogeneous Generalized Master Equations (GMEs),
governing the evolution in time of equilibrium two-time correlation functions
for dynamic variables of a subsystem of $s$ particles ($s<N$) selected from
$N>>1$ particles of a classical many-body system, are obtained These
time-convolution and time-convolutionless GMEs differ from the known GMEs
(e.g. Nakajima-Zwanzig GME) by absence of inhomogeneous terms containing
correlations between all $N$ particles at the initial moment of time and
preventing the closed description of $s$-particles subsystem evolution. Closed
homogeneous GMEs describing the subdynamics of fluctuations are obtained by
applying a special projection operator to the Liouville equation governing the
dynamics of $N$-particle system. In the linear approximation in the particles'
density, the linear Generalized Boltzmann equation accounting for initial
correlations and valid at all timescales is obtained \ This equation for a
weak inter-particle interaction converts into the generalized linear Landau
equation in which the initial correlations are also accounted for. Connection
of these equations to the nonlinear Boltzmann and Landau equations are discussed.

\end{abstract}
\maketitle

\section{Introduction}

One of the long-standing problems of statistical physics of $N$-particle
($N>>1$) system remains the derivation of the closed kinetic equations for
$s$-particle ($s<N$) distribution functions sufficient for calculation of the
measurable values characterizing a non-equilibrium sate of that many-particle
system. The natural starting point is the Liouville (classical system) or
von-Neumann (quantum system) linear equation for an $N$-particfle distribution
function or statistical operator. In the reduced description method leading to
the BBGKY chain, the closed Boltzmann kinetic equation can be obtained by
employing the Boltzmann "molecular chaos" approximation at any time moment
(beginning \ from the initial state) or more sophisticated Bogoliubov
principle of weakening of initial correlations \cite{Bogoliubov}. In the
latter case, in the first approximation in particles' density, the nonlinear
Boltzmann equation follows from the BBGKY chain on a large (kinetic)
timescale. Note, that nonlinearity of the Boltzmann equation, obtained from
the linear Liouville (or von-Neumann) equation, is a consequence of the above
mentioned approximations. Lanford's derivation of the Boltzmann equation
(however only on a small timescale) \cite{Lanford (1975)} seems to be the most
relevant result in the mathematical foundation of the kinetic theory.

For the case of small inter-particle interaction, the nonlinear Landau
equation follows from the Boltzmann equation (see, e.g., \cite{Balescu}). But,
strictly speaking, the Landau equation should be derived from the particles
system dynamics (the Liouville equation) in the weak-coupling limit. The
partial result (for the short timescale) was obtained in \cite{Bobylev (2013)}.

In the projection operators approach leading to the Generalized Master
Equations (GMEs) (see, e.g., \cite{Breuer}), in order to obtain the completely
closed (homogeneous) linear equation for the reduced $s$-particle ($s<N$)
distribution function (statistical operator), the undesired inhomogeneous term
(a source) containing all $N$-particle initial correlations should be
disregarded (which is incorrect in principle \cite{Van Kampen}).

These procedures allowing for obtaining the desired closed kinetic equations
are not completely satisfactory: They implies, e.g., the "propagation of
chaos" in time hypothesis for the factorized initial state, the general proof
of which is still lacking (see, e.g. \cite{Kac}), and do not allow for
considering the evolution process on any timescale for arbitrary initial
state. The natural desire then arises to abandon the "molecular chaos" (or
other mentioned assumption) and include initial correlations into
consideration. This can be effectively done, e.g., by deriving from the
Liouville (von Neumann) equation the completely closed (homogeneous with no
source) evolution equations valid on any timescale. For arbitrary initial
conditions it has been attempted in \cite{Los JPA, Los JSP, Los Evolution
equations, Los Theor. Math. Phys}.

On the other hand, the equilibrium state for the full system provides the
natural initial state for the system under consideration \cite{Van Kampen}.
Moreover, this case provides an additional opportunity for including initial
correlations into consideration. For quantum system, the closed homogeneous
linear evolution equations accounting for initial correlations was obtained in
\cite{Los 2017, Los 2018} with application to polaron mobility problem.

It turns out, that the initial equilibrium state in the classical physics case
is even more favorable, than the case of quantum physics, for realization of
the program with no "molecular chaos" (factorized initial state) assumption.
In this paper we show that there is a special projection operator selecting
the relevant part of the equilibrium two-time $s$-particle correlation
function for the $N$-particle ($N>s$) system of classical particles which
obeys the exact time-convolution (TC) or time-convolutionless (TCL)
homogeneous GMEs. Thus, it is shown that there is a subdynamics in the
subspace of $s$ particles, and, therefore, the evolution of the correlation
function (thermal fluctuations) is completely closed (no undesirable terms
defined on the full phase space of $N$ particles). The initial correlations
are "hidden" in the projection operator which can be expanded in the particles
density $n$ or in a small inter-particle interaction. In the first case, the
linear generalized Boltzmann equation accounting for initial correlations and
valid at all timescales follows from the obtained homogeneous TC GME in the
first in $n$ order expansion of the kernel governing the evolution of a
one-particle correlation function. We show how this equation leads to the
known linear Boltzmann equation with additional term related to initial
correlations. In the case of small inter-particle interaction, we obtain from
this equation the linear equation for one-particle correlation function and
discuss its connection to the Landau equation.

\section{Projection operator formalism for s-particle equilibrium correlation
function}

We consider an $N$-particle ($N>>1$) system of interacting classical
particles. Let us select the subsystem $s$, i.e., the complex of $s$ ($s<N$)
particles ($s$-complex), which interacts with the environment $\Sigma$ of
remaining $N-s$ particles. Note, that the particles, making up a subsystem,
can be of different from the environment particles sort. Then, we assume that
the Hamilton function of the full system can be presented as
\begin{equation}
H=H_{s}+H_{\Sigma}+\widetilde{H}_{s\Sigma},\label{1}%
\end{equation}
where $H_{s}$, $H_{\Sigma}$ and $\widetilde{H}_{s\Sigma}$ are the Hamilton
functions of the subsystem $s$, environment $\Sigma$ and the
subsystem-environment interaction $\widetilde{H}_{s\Sigma}$, respectively.
More specific form of these functions will be considered later.

We consider a two-time equilibrium correlation function for subsystem's
dynamic functions $A_{s}$ and $B_{s}$, which depend on the set of variables
characterizing the subsystem, i.e on $x_{i}=(\mathbf{r}_{i},\mathbf{p}_{i})$,
$i=1,2,...,s$, where $x_{i}$ is the coordinate of the $i$-th particle in the
phase space. The time dependence of dynamic functions is given by
$A_{s}(t)=\exp(Lt)A_{s}(0)$, $B_{s}(t)=\exp(Lt)B_{s}(0)$, where $L$ is the
Liouville operator $L=L_{s}+L_{\Sigma}+\widetilde{L}_{s\Sigma}$ related to the
Hamilton function (\ref{1}) and defined by the Poisson bracket. Thus, we
consider the correlation function
\begin{align}
\varphi_{AB}(t)  & =<A_{s}(t)B_{s}(0)>=<A_{s}(0)B_{s}(-t)>=\int...\int
dx^{s}A_{s}(0)[\int...\int dx^{\Sigma}G_{N}(t,\beta)]B_{s}(0),\nonumber\\
G_{N}(t,\beta)  & =\rho(\beta)\exp(-Lt),dx^{s}=dx_{1}...dx_{s},dx^{\Sigma
}=dx_{s+1}...dx_{N}.\label{2}%
\end{align}
Here,
\begin{align}
\rho(\beta)  & =Z^{-1}\exp(-\beta H),Z=\int...\int dx^{N}\exp(-\beta
H),\nonumber\\
LC_{N}  & =\{H,C_{N}\}_{p}\nonumber\\
& =%
%TCIMACRO{\dsum \limits_{i=1}^{N}}%
%BeginExpansion
{\displaystyle\sum\limits_{i=1}^{N}}
%EndExpansion
\left[  \frac{\partial C_{N}}{\partial\mathbf{r}_{i}}\frac{\partial
H}{\partial\mathbf{p}_{i}}-\frac{\partial C_{N}}{\partial\mathbf{p}_{i}}%
\frac{\partial H}{\partial\mathbf{r}_{i}}\right]  ,\nonumber\\
dx^{N}  & =dx^{s}dx^{\Sigma},\label{3}%
\end{align}
$\{H,C_{N}\}_{p}$ is the Poisson bracket, $C_{N}$ is some dynamic function
defined on the full phase space of the $N$-particle system under
consideration. We see, that the dynamics of correlation function (\ref{2}) is
defined by function $G_{N}(t,\beta)$ which depends on the whole set of
variables $x_{1},...,x_{N}$ and obeys the equation%
\begin{equation}
\frac{\partial}{\partial t}G_{N}(t,\beta)=-LG_{N}(t,\beta).\label{4}%
\end{equation}
The formal solution to Eq. (\ref{4}) is
\begin{align}
G_{N}(t,\beta)  & =U(t,0)G_{N}(0,\beta),\nonumber\\
U(t,0)  & =\exp(-Lt).\label{4'}%
\end{align}
However, it is seen from the definition (\ref{2}), that dynamics of the
subsystem fluctuations is governed by function dependent on much smaller
number of variables $x_{1},...,x_{s}$ than the whole set of $N$ variables
$x_{1},...,x_{N}$, i.e., by function $G_{N}(t,\beta)$ integrated over the
environment variables $x_{s+1},..,x_{N}$
\begin{equation}
F_{s}(t,\beta)=\int...\int dx^{\Sigma}G_{N}(t,\beta).\label{4a}%
\end{equation}
\qquad

In order to obtain the equation for the reduced function $F_{s}(t,\beta)$
(\ref{4a}), it is convenient to employ the projection operator technique
\cite{Nakajima (1958)}, \cite{Zwanzig (1960)}, \cite{Prigogine (1962)} and to
break $G_{N}(t,\beta)$ by some projection operators $P$ and $Q=1-P$ (with the
properties $P^{2}=P$, $Q^{2}=Q$, $P+Q=1$, $PQ=0$) into the relevant
$R_{N}(t,\beta)$ and irrelevant $I_{N}(t,\beta)$ parts
\begin{subequations}
\begin{align}
G_{N}(t,\beta) &  =R_{N}(t,\beta)+I_{N}(t,\beta),\nonumber\\
R_{N}(t,\beta) &  =PG_{N}(t,\beta),I_{N}(t,\beta)=QG_{N}(t,\beta
)=G_{N}(t,\beta)-R_{N}(t,\beta).\label{4b}%
\end{align}
We note, that the relevant and irrelevant parts generally depend on
coordinates and momenta of all $N$ particles in contrast to the reduced
function $F_{s}(t,\beta)$. The relevant part $R_{N}(t,\beta)$ is conveniently
defined in such a way that it comprises the reduced function of interest
$F_{s}(t,\beta)$ as a multiplier. Thus, we consider the projection operators
of the form
\end{subequations}
\begin{equation}
P=\Phi_{s\Sigma}\int...\int dx^{\Sigma},\label{5}%
\end{equation}
where the function $\Phi_{s\Sigma}$ generally depends on the coordinates of a
subsystem and an environment and normalized as
\begin{equation}
\int...\int\Phi_{s\Sigma}dx^{\Sigma}=1.\label{6}%
\end{equation}
Then, it is easily seen that for projectors given by (\ref{5}) and (\ref{6}),
the correlation function (\ref{2}) is completely defined by the relevant part
of $G_{N}(t,\beta)$
\begin{equation}
\varphi_{AB}(t)=\int...\int dx^{s}\int...\int dx^{\Sigma}A_{s}(0)R_{N}%
(t,\beta)B_{s}(0).\label{6a}%
\end{equation}
If, e.g.,
\begin{equation}
\Phi_{s\Sigma}=\rho_{\Sigma}=Z_{\Sigma}^{-1}\exp(-\beta H_{\Sigma}),Z_{\Sigma
}=\int...\int dx^{\Sigma}\exp(-\beta H_{\Sigma}),\label{7}%
\end{equation}
then we have the "standard" projectors (see, e.g., \cite{Breuer})
\begin{equation}
P=P_{\Sigma}=\rho_{\Sigma}\int...\int dx^{\Sigma},Q_{\Sigma}=1-P_{\Sigma
}\label{8}%
\end{equation}
conventionally used for such types of problems (interaction of a subsystem
with a reservoir).

By application of operators (\ref{8}) to equation (\ref{4}), we obtain the
equations for the relevant and irrelevant parts of $G_{N}(t,\beta)$
\begin{align}
\frac{\partial}{\partial t}R_{N}(t,\beta) &  =-P_{\Sigma}L[R_{N}%
(t,\beta)+I_{N}(t,\beta)],\nonumber\\
\frac{\partial}{\partial t}I_{N}(t,\beta) &  =-Q_{\Sigma}L[R_{N}%
(t,\beta)+I_{N}(t,\beta)],\label{9}%
\end{align}
where now
\begin{equation}
R_{N}(t,\beta)=P_{\Sigma}G_{N}(t,\beta)=G_{N}^{r}(t,\beta)=\rho_{\Sigma}%
F_{s}(t,\beta),I_{N}(t,\beta)=G_{N}(t,\beta)-\rho_{\Sigma}F_{s}(t,\beta
)=G_{N}^{i}(t,\beta).\label{9a}%
\end{equation}
Finding $G_{N}^{i}(t,\beta)$ from the second equation (\ref{9}) as a function
of $G_{N}^{r}(\tau,\beta)$ and $G_{N}^{i}(0,\beta)$ and inserting it in the
first equation (\ref{9}), we arrive at the the conventional exact
time-convolution generalized master equation (TC-GME) known as the
Nakajima-Zwanzig equation for the relevant part of function $G_{N}^{r}%
(t,\beta)$ \cite{Nakajima (1958)}, \cite{Zwanzig (1960)}
\begin{align}
\frac{\partial}{\partial t}G_{N}^{r}(t,\beta) &  =-P_{\Sigma}LG_{N}%
^{r}(t,\beta)+\int\limits_{0}^{t}P_{\Sigma}LU_{Q_{\Sigma}}(t,\tau)Q_{\Sigma
}LG_{N}^{r}(\tau,\beta)d\tau\nonumber\\
&  -P_{\Sigma}LU_{Q_{\Sigma}}(t,0)G_{N}^{i}(0,\beta),\nonumber\\
U_{Q_{\Sigma}}(t,\tau) &  =\exp[-Q_{\Sigma}LQ_{\Sigma}(t-\tau)].\label{10}%
\end{align}
Equation (\ref{10}), which, in fact, gives the equation for the reduced
function $F_{s}(t,\beta)$, is quite general and formally closed. Serving as a
basis for many applications \cite{Breuer}, this equation, nevertheless,
contains the undesirable and non-negligible inhomogeneous initial condition
term (the last term in the right hand side of (\ref{10}))%
\begin{align}
G_{N}^{i}(0,\beta)  & =G_{N}(0,\beta)-P_{\Sigma}G_{N}(0,\beta)=\rho
(\beta)-\rho_{s}\rho_{\Sigma},\nonumber\\
\rho_{s}  & =\int...\int dx^{\Sigma}\rho(\beta)\label{11}%
\end{align}
(see (\ref{2}) and (\ref{8})). This term is not equal to zero due to initial
(at $t=0$) correlations ($\rho(\beta)\neq\rho_{s}\rho_{\Sigma}$). Therefore,
Eq. (\ref{10}) does not provide for a complete reduced description of a
multiparticle system in terms of the relevant (reduced) function
$F_{s}(t,\beta)$. Applying Bogoliubov's principle of weakening of initial
correlations, allowing to eliminate the influence of initial correlations on
the large enough time scale $t\gg t_{cor}$ ($t_{cor}$ is the time for damping
of initial correlations) or using a factorized initial condition, when
$\rho_{\text{ }}(\beta)=\rho_{s}\rho_{\Sigma}$, one can achieve the desirable
goal and obtain the homogeneous GME for $F_{s}(t,\beta)$, i.e. Eq. (\ref{10})
with no initial condition term. However, obtained in such a way homogeneous
GME is either approximate and valid only on a large enough time scale (when
all initial correlations vanish) or applicable only for a rather artificial
(actually unreal, as pointed in \cite{Van Kampen}) initial conditions (no
correlations at an initial instant of time). In addition, Eq. (\ref{10}) poses
the problem to deal with due to its time-nonlocality. However, it is possible
to obtain the time-local equation for the relevant part $G_{N}^{r}(t,\beta)$
\cite{Breuer, Shibata1 (1977), Shibata2 (1980)} which also contains the
inhomogeneous source term.

\section{Completely closed (homogeneous) GMEs for s-particle correlation
function}

Let us now introduce the following projection operators $P_{s}$ and $Q_{s}$
\begin{align}
P &  =P_{s}=\rho_{\Sigma}^{s}\int...\int dx^{\Sigma},Q=Q_{s}=1-P_{s}%
,\nonumber\\
\rho_{\Sigma}^{s} &  =\frac{1}{Z_{\Sigma}^{s}}\exp[-\beta(H_{\Sigma
}+\widetilde{H}_{s\Sigma})],\nonumber\\
Z_{\Sigma}^{s} &  =\int...\int dx^{\Sigma}\exp[-\beta(H_{\Sigma}+\widetilde
{H}_{s\Sigma})].\label{16}%
\end{align}
It is not difficult to see that $P_{s}^{2}=P_{s}$, $Q_{s}^{2}=Q_{s}$,
$P_{s}Q_{s}=0$. Then, we can divide $G_{N}(t,\beta)$ into the relevant
$g_{N}^{r}(t,\beta)$ and irrelevant $g_{N}^{i}(t,\beta)$\ components as\ %

\begin{align}
G_{N}(t,\beta) &  =g_{N}^{r}(t,\beta)+g_{N}^{i}(t,\beta),\nonumber\\
g_{N}^{r}(t,\beta) &  =P_{s}G_{N}(t,\beta)=\rho_{\Sigma}^{s}F_{S}%
(t,\beta),\nonumber\\
g_{N}^{i}(t,\beta) &  =Q_{s}G_{N}(t,\beta)=G_{N}(t,\beta)-\rho_{\Sigma}%
^{s}F_{S}(t,\beta).\label{17}%
\end{align}
It is not difficult to see that the dynamics of the correlation function
(\ref{2}) is completely defined by the relevant part $g_{N}^{r}(t,\beta)$ of
$G_{N}(t,\beta)$, i.e.,%
\begin{equation}
\varphi_{AB}(t)=\int...\int dx^{s}A_{s}(0)F_{s}(t,\beta)B_{s}(0)=\int...\int
dx^{N}A_{s}(0)g_{N}^{r}(t,\beta)B_{s}(0).\label{18}%
\end{equation}
The projection operator $P_{s}$ (\ref{16}) has an interesting property,
namely,%
\begin{equation}
P_{s}\rho(\beta)=\rho(\beta),Q_{s}G_{N}(0,\beta)=0.\label{19}%
\end{equation}

Thus, \bigskip by applying the introduced projection operators $P_{s}$ and
$Q_{s}$ (\ref{16}) to equation (\ref{4}), we arrive at the following exact
homogeneous time-convolution GME (compare with (\ref{10}))%

\begin{align}
\frac{\partial}{\partial t}g_{N}^{r}(t,\beta) &  =-P_{s}Lg_{N}^{r}%
(t,\beta)+\int\limits_{0}^{t}P_{s}LU_{Q_{s}}(t,\tau)Q_{s}Lg_{N}^{r}(\tau
,\beta)d\tau,\nonumber\\
U_{Q_{s}}(t,\tau) &  =\exp[-Q_{s}L(t-\tau)].\label{20}%
\end{align}

Equation (\ref{20}) is the completely closed equation for the relevant part of
the correlation function that we are looking for. It shows, that in the
considered case, the dynamics of correlation function (\ref{2}) can be exactly
projected on the dynamics within its relevant subspace. It follows, that the
correlation of fluctuations of the selected complex of $s$ particles can be
described by the \textbf{linear} equation in the subspace of the corresponding
coordinates $x_{i}=(\mathbf{r}_{i},\mathbf{p}_{i})$ ($i=1,...,s$). To make it
more clear, we rewrite Eq. (\ref{20}) as the equation for an $s$-particle
function $F_{s}(t,\beta)$ (\ref{4a}) ) governing the subsystem's fluctuations
in time (see also (\ref{2}) and (\ref{17}))
\begin{equation}
\frac{\partial}{\partial t}F_{s}(t,\beta)=-\left[  \int...%
%TCIMACRO{\dint }%
%BeginExpansion
{\displaystyle\int}
%EndExpansion
dx^{\Sigma}L\rho_{\Sigma}^{s}\right]  F_{s}(t,\beta)+\left[  \int...%
%TCIMACRO{\dint }%
%BeginExpansion
{\displaystyle\int}
%EndExpansion
dx^{\Sigma}L\int\limits_{0}^{t}d\tau U_{Q_{s}}(t,\tau)Q_{s}L\rho_{\Sigma}%
^{s}\right]  F_{s}(\tau,\beta).\label{21}%
\end{equation}

Generally, the evolution equation (\ref{20}) poses some problem to deal with
due to its time-nonlocality. It is possible, however, to obtain the exact
homogeneous time-local equation for the relevant part of the correlation
function. The idea is to take advantage of the evolution of $G_{N}(t,\beta)$,
defined by (\ref{4'}), which leads to the relation
\begin{align}
G_{N}(\tau,\beta)  & =U^{-1}(t,\tau)G_{N}(t,\beta),\nonumber\\
U^{-1}(t,\tau)  & =\exp[L(t-\tau)].\label{22}%
\end{align}
Using (\ref{22}) and the conventional projection operator (\ref{8}), the well
known time-convolutionless equation for the relevant part $G_{N}^{r}(t,\beta)$
of $G_{N}(t,\beta)$, which contains the undesirable inhomogeneous term
(\ref{11}) comprising the initial correlations, can be obtained.(see
\cite{Breuer, Shibata1 (1977), Shibata2 (1980)}).

We will show now, that the use of the projector (\ref{16}) instead of
(\ref{8}) leads to the completely closed homogeneous time-convolutionless GME
for the relevant part of the correlation function. We will briefly conduct the
derivation which is rather a standard one. First, we apply the projector
(\ref{16}) to (\ref{22}) and obtain additional equation connecting the
relevant and irrelevant parts of $G_{N}(t,\beta)$
\begin{equation}
g_{N}^{r}(\tau,\beta)=P_{s}U^{-1}(t,\tau)[g_{N}^{r}(t,\beta)+g_{N}^{i}%
(t,\beta)].\label{23}%
\end{equation}
We also have the equation for the irrelevant part $g_{N}^{i}(t,\beta)$ which
follows from the solution of the second equation (\ref{9}) with the projection
operator (\ref{16})
\begin{equation}
g_{N}^{i}(t,\beta)=-%
%TCIMACRO{\dint \limits_{0}^{t}}%
%BeginExpansion
{\displaystyle\int\limits_{0}^{t}}
%EndExpansion
U_{Q_{s}}(t,\tau)Q_{s}Lg_{N}^{r}(\tau,\beta)d\tau,\label{24}%
\end{equation}
where $U_{Q_{s}}(t,\tau)$ is given by (\ref{20}) and the property (\ref{19})
was used. From two equations (\ref{23}) and (\ref{24}) one finds that
\begin{align}
g_{N}^{i}(t,\beta) &  =[1-\alpha(t)]^{-1}\alpha(t)g_{N}^{r}(t,\beta
),\nonumber\\
\alpha(t) &  =-%
%TCIMACRO{\dint \limits_{0}^{t}}%
%BeginExpansion
{\displaystyle\int\limits_{0}^{t}}
%EndExpansion
U_{Q_{s}}(t,\tau)Q_{s}LP_{s}U^{-1}(t,\tau)d\tau.\label{25}%
\end{align}
Substituting $g_{N}^{i}(t,\beta)$ (\ref{25}) into the \ projected by $P_{s}$
equation (\ref{4})%
\begin{equation}
\frac{\partial}{\partial t}g_{s}^{r}(t,\beta)=-P_{s}L[g_{N}^{r}(t,\beta
)+g_{N}^{i}(t,\beta)],\label{26}%
\end{equation}
we finally obtain%
\begin{equation}
\frac{\partial}{\partial t}g_{N}^{r}(t,\beta)=-P_{s}L[1-\alpha(t)]^{-1}%
g_{N}^{r}(t,\beta).\label{27}%
\end{equation}
\qquad If it is possible to expand the operator $[1-\alpha(t)]^{-1}$ into the
series in $\alpha(t)$, then the first two terms of this expansion results in
the following time-local equation (compare with (\ref{20}))%
\begin{equation}
\frac{\partial}{\partial t}g_{N}^{r}(t,\beta)=-P_{s}Lg_{N}^{r}(t,\beta)+P_{s}L%
%TCIMACRO{\dint \limits_{0}^{t}}%
%BeginExpansion
{\displaystyle\int\limits_{0}^{t}}
%EndExpansion
d\tau U_{Q_{s}}(t,\tau)Q_{s}LP_{s}U^{-1}(t,\tau)g_{N}^{r}(t,\beta).\label{28}%
\end{equation}

Equations (\ref{20}) and (\ref{27}) present the main results of this section.
They show that the projector (\ref{16}) allows for selecting the relevant part
$g_{N}^{r}(t,\beta)$ of the multiparticle function $G_{N}(t,\beta)$ governing
the dynamics of correlation function function (\ref{2}) which satisfies the
completely closed linear time-convolution and time-convolutionless equations.
They, in fact, describe the evolution of the $s$-particles marginals
(\ref{4a}) on the arbitrary timescale. Thus, one remains in the scope of the
linear evolution given by the Liouville equation (\ref{4}) but should pay for
this simplification by accounting for initial correlations, which are
conveniently ignored. It is also worth noting that the developed formalism
only works in the framework of classical physics (when the terms of the
Hamilton function (\ref{1}) commutes with each other). For quantum physics a
different approach is needed (see \cite{Los 2017, Los 2018}).

\section{Equations for a more specific case}

Let us specify the Hamilton function $H$ (\ref{1}) for the case of the
identical particles with the two-body interparticle interaction $V_{ij}$ as%
\begin{align}
H &  =H_{s}+H_{\Sigma}+\widetilde{H}_{s\Sigma},\nonumber\\
H_{s} &  =\sum\limits_{i=1}^{s}\frac{\mathbf{p}_{i}^{2}}{2m}+\sum
\limits_{1\leq i<j\leq s}V_{ij}(\left\vert \mathbf{r}_{i}-\mathbf{r}%
_{j}\right\vert )+<H_{s\Sigma}>_{\Sigma},\nonumber\\
H_{\Sigma} &  =%
%TCIMACRO{\dsum \limits_{i=s+1}^{N}}%
%BeginExpansion
{\displaystyle\sum\limits_{i=s+1}^{N}}
%EndExpansion
\frac{\mathbf{p}_{i}^{2}}{2m}+\sum\limits_{s+1\leq i<j\leq N}V_{ij}(\left\vert
\mathbf{r}_{i}-\mathbf{r}_{j}\right\vert ),\nonumber\\
\widetilde{H}_{s\Sigma} &  =H_{s\Sigma}-<H_{s\Sigma}>_{\Sigma},H_{s\Sigma
}=\sum\limits_{i=1}^{s}\sum\limits_{j=s+1}^{N}V_{ij}(\left\vert \mathbf{r}%
_{i}-\mathbf{r}_{j}\right\vert ).\label{29}%
\end{align}
Here, for convenience, we introduce the energy of the mean field $<H_{s\Sigma
}>_{\Sigma}$acting on the $s$-complex by the "equilibrium" environment%
\begin{equation}
<H_{s\Sigma}>_{\Sigma}=\int...%
%TCIMACRO{\dint }%
%BeginExpansion
{\displaystyle\int}
%EndExpansion
dx^{\Sigma}\rho_{\Sigma}H_{s\Sigma},\label{30}%
\end{equation}
where $\rho_{\Sigma}$ is given by (\ref{7}). Note, that $<H_{s\Sigma}%
>_{\Sigma}$depends only on the coordinates of $s$ selected particles
$\mathbf{r}_{i}$ ($i=1,...,s$). For a space-homogeneous case, this mean field
does not depend on $\mathbf{r}_{i}$ ($i=1,...,s$).

The corresponding to (\ref{29}) Liouville operator $L$ is%
\begin{align}
L &  =L_{s}+L_{\Sigma}+\widetilde{L}_{s\Sigma},\nonumber\\
L_{s} &  =%
%TCIMACRO{\dsum \limits_{i=1}^{s}}%
%BeginExpansion
{\displaystyle\sum\limits_{i=1}^{s}}
%EndExpansion
[\mathbf{v}_{i}\mathbf{\nabla}_{i}-(\mathbf{\nabla}_{i}<H_{s\Sigma}>_{\Sigma
})\frac{\partial}{\partial\mathbf{p}_{i}}]-\sum\limits_{1\leq i<j\leq
s}(\mathbf{\nabla}_{i}V_{ij})\cdot(\frac{\partial}{\partial\mathbf{p}_{i}%
}-\frac{\partial}{\partial\mathbf{p}_{j}}),\nonumber\\
L_{\Sigma} &  =%
%TCIMACRO{\dsum \limits_{i=s+1}^{N}}%
%BeginExpansion
{\displaystyle\sum\limits_{i=s+1}^{N}}
%EndExpansion
\mathbf{v}_{i}\mathbf{\nabla}_{i}-\sum\limits_{s+1\leq i<j\leq N}%
(\mathbf{\nabla}_{i}V_{ij})\cdot(\frac{\partial}{\partial\mathbf{p}_{i}}%
-\frac{\partial}{\partial\mathbf{p}_{j}}),\nonumber\\
\widetilde{L}_{s\Sigma} &  =-\sum\limits_{i=1}^{s}\sum\limits_{j=s+1}%
^{N}(\mathbf{\nabla}_{i}V_{ij})\cdot(\frac{\partial}{\partial\mathbf{p}_{i}%
}-\frac{\partial}{\partial\mathbf{p}_{j}})+%
%TCIMACRO{\dsum \limits_{i=1}^{s}}%
%BeginExpansion
{\displaystyle\sum\limits_{i=1}^{s}}
%EndExpansion
(\mathbf{\nabla}_{i}<H_{s\Sigma}>_{\Sigma})\frac{\partial}{\partial
\mathbf{p}_{i}},\nonumber\\
\mathbf{v}_{i} &  =\mathbf{p}_{i}/m,\mathbf{\nabla}_{i}=\frac{\partial
}{\partial\mathbf{r}_{i}},V_{ij}=V_{ij}(\left\vert \mathbf{r}_{i}%
-\mathbf{r}_{j}\right\vert ).\label{31}%
\end{align}

Equations (\ref{20}) and (\ref{27}) can be rewritten in the simplified form if
we take into account the following operators properties%
\begin{align}
P_{s}L_{s} &  =P_{s}L_{s}P_{s}=\rho_{\Sigma}^{s}L_{s}\int...%
%TCIMACRO{\dint }%
%BeginExpansion
{\displaystyle\int}
%EndExpansion
dx^{\Sigma},P_{s}L_{s}Q_{s}=0,\nonumber\\
Q_{s}L_{s}P_{s} &  =L_{s}P_{s}-\rho_{\Sigma}^{s}L_{s}\int...%
%TCIMACRO{\dint }%
%BeginExpansion
{\displaystyle\int}
%EndExpansion
dx^{\Sigma},\nonumber\\
P_{s}L_{\Sigma} &  =0,P_{s}L_{\Sigma}Q_{s}=0,Q_{s}L_{\Sigma}P_{s}=L_{\Sigma
}P_{s},\nonumber\\
P_{s}\widetilde{L}_{s\Sigma}P_{s} &  =%
%TCIMACRO{\dsum \limits_{i=1}^{s}}%
%BeginExpansion
{\displaystyle\sum\limits_{i=1}^{s}}
%EndExpansion
[<\mathbf{F}_{i}>_{\Sigma}^{s}-<\mathbf{F}_{i}>_{\Sigma}]\frac{\partial
}{\partial\mathbf{p}_{i}}P_{s},\label{32}%
\end{align}
where%
\begin{equation}
\mathbf{F}_{i}=-%
%TCIMACRO{\dsum \limits_{j=s+1}^{N}}%
%BeginExpansion
{\displaystyle\sum\limits_{j=s+1}^{N}}
%EndExpansion
(\mathbf{\nabla}_{i}V_{ij}),<...>_{\Sigma}^{s}=\int...%
%TCIMACRO{\dint }%
%BeginExpansion
{\displaystyle\int}
%EndExpansion
dx^{\Sigma}(...\rho_{\Sigma}^{s}),\label{33}%
\end{equation}
i.e., $\mathbf{F}_{i}$ is the force acting on the $i$-particle ($i=1,...,s$)
from the the "environment" of $N-s$ particles. Here and further on, we use, as
usual, that all functions $\Phi(x_{1},\ldots,x_{N};t)$, defined on the phase
space, and their derivatives vanish at the boundaries of the configurational
space and at $\mathbf{p}_{i}=\pm\infty$.

Now Eq. (\ref{20}) can be presented as%

\begin{equation}
\frac{\partial}{\partial t}F_{s}(t,\beta)=-\left[  L_{s}+\sum\limits_{i=1}%
^{s}(<\mathbf{F}_{i}>_{\Sigma}^{s}-<\mathbf{F}_{i}>_{\Sigma})\frac{\partial
}{\partial\mathbf{p}_{i}}\right]  F_{s}(t,\beta)+C(t,\beta),\label{33a}%
\end{equation}
where the collision term $C(t,\beta)$ is defined by%
\begin{align}
C(t,\beta)  & =\int\limits_{0}^{t}d\tau\int...\int dx^{\Sigma}\widetilde
{L}_{s\Sigma}U_{Q_{s}}(\tau,0)[\widetilde{L}_{s\Sigma}\rho_{\Sigma}^{s}%
+(L_{s}\rho_{\Sigma}^{s})+L_{\Sigma}\rho_{\Sigma}^{s}\nonumber\\
-\sum\limits_{i=1}^{s}(  & <\mathbf{F}_{i}>_{\Sigma}^{s}-<\mathbf{F}%
_{i}>_{\Sigma})\rho_{\Sigma}^{s}\frac{\partial}{\partial\mathbf{p}_{i}}%
]F_{s}(t-\tau,\beta).\label{33b}%
\end{align}
and $(L_{s}\rho_{\Sigma}^{s})$ means that $L_{s}$ acts only on $\rho_{\Sigma
}^{s}$. Note that if we use in (\ref{33a}) and (\ref{33b}) the standard
projector (\ref{8}), i.e., substitute $\rho_{\Sigma}^{s}$ with $\rho_{\Sigma}%
$, then Eq. (\ref{33a}) will acquire the "standard" form
\begin{equation}
\frac{\partial}{\partial t}F_{s}(t,\beta)=-L_{s}F_{s}(t,\beta)+\int
\limits_{0}^{t}d\tau\int...\int dx^{\Sigma}\widetilde{L}_{s\Sigma}U_{Q_{s}%
}(\tau,0)\widetilde{L}_{s\Sigma}\rho_{\Sigma}F_{s}(t-\tau,\beta)\label{33c}%
\end{equation}
if we neglect in (\ref{10}) the inhomogeneous source term. Thus, we see that
the extra terms in Eq. (\ref{33a}) are due to initial correlations which are
"hidden" in the projector (\ref{16}).

\section{Evolution equation for one-particle correlation function in the first
approximation in the particles density}

For what follows, we consider the equation for $F_{1}(t,\beta)=F_{1}%
(\mathbf{r}_{1},\mathbf{p}_{1};t,\beta)$. One can see that for $s=1$ the
Hamilton function (\ref{29}) and the Liouville operator (\ref{31}) have more
simple form. In order to expand the kernel of Eq. (\ref{33a}) in the density
of particles $n=N/V$ ($V$ is the system's volume), we need the expansion for
the distribution function $\rho_{\Sigma}^{1}$. In order to do that, it is
convenient to express $\exp(-\beta H_{\Sigma})$ and $\exp(-\beta H_{1\Sigma})$
in terms of the Mayer functions $f_{ij}$ \cite{Mayer}. Then,
\begin{align}
\rho_{\Sigma}^{1}  & =\frac{\exp(-\beta H_{\Sigma})\exp(-\beta H_{1\Sigma}%
)}{\int...\int dx^{\Sigma}\exp(-\beta H_{\Sigma})\exp(-\beta H_{1\Sigma}%
)}\nonumber\\
& =\rho_{\Sigma}(\mathbf{p})\frac{\prod\limits_{2\leq i<j\leq N}%
(1+f_{ij})\prod\limits_{j=2}^{N}(1+f_{1j})}{\int...\int d\mathbf{r}^{\Sigma
}\prod\limits_{2\leq i<j\leq N}(1+f_{ij})\prod\limits_{j=2}^{N}(1+f_{1j}%
)},d\mathbf{r}^{\Sigma}=d\mathbf{r}_{2}...d\mathbf{r}_{N}\label{33d}%
\end{align}
where%
\begin{align}
f_{ij}  & =e^{-\beta V_{ij}}-1,V_{ij}=V(\left\vert \mathbf{r}_{i}%
-\mathbf{r}_{j}\right\vert ),\nonumber\\
\rho_{\Sigma}(\mathbf{p})  & =\exp[-\beta H_{\Sigma}(\mathbf{p})]/\int...\int
d\mathbf{p}^{\Sigma}\exp[-\beta H_{\Sigma}(\mathbf{p})],\nonumber\\
H_{\Sigma}(\mathbf{p})  & =\sum\limits_{j-2}^{N}p_{j}^{2}/2m,d\mathbf{p}%
^{\Sigma}=d\mathbf{p}_{2}...d\mathbf{p}_{N},\label{33e}%
\end{align}
and we used that $\exp(\beta<H_{1\Sigma}>_{\Sigma})$ does not depend on the
"environment" $\Sigma$ variables and is cancelled out of $\rho_{\Sigma}^{1}$.

Let us consider the denominator of $\rho_{\Sigma}^{1}$
\begin{align}
& \int...\int d\mathbf{r}^{\Sigma}\prod\limits_{2\leq i<j\leq N}%
(1+f_{ij})\prod\limits_{j=2}^{N}(1+f_{1j})\nonumber\\
& =\int d\mathbf{r}_{2}...\int d\mathbf{r}_{N}(1+f_{23}+f_{24}+f_{34}%
+...+f_{N-1,N}+f_{23}f_{24}+...)\nonumber\\
& \times(1+f_{12}+f_{13}+...+f_{12}f_{13}+...)\nonumber\\
& =V^{N-1}+(N-1)V^{N-2}\int f_{12}d\mathbf{r}_{2}+N(N-1)V^{N-3}\int
d\mathbf{r}_{2}\int d\mathbf{r}_{3}(f_{23}+f_{12}f_{13})+...\nonumber\\
& =V^{N-1}[1+n\int d\mathbf{r}_{2}f_{12}+n^{2}\int d\mathbf{r}_{2}\int
d\mathbf{r}_{3}(f_{23}+f_{12}f_{13})+...],N>>1.\label{33f}%
\end{align}
One can see that the terms with one integration over the particle coordinate
is proportional to $n$, while the terms with integration over coordinates of
two, three \ and more particles are proportional to $n^{2}$, $n^{3}$ and
higher powers of $n$, respectively. In what follows, we will restrict
ourselves to the linear in $n$ approximation to the kernel of Eq. (\ref{33a}),
and, therefore, only the terms with one integration over the particle
coordinate should be taken into consideration. The products of terms with one
integration over the particle coordinate will also be (naturally) disregarded.
Thus, in the linear in $n$ approximation
\begin{equation}
\rho_{\Sigma}^{1}=\frac{1}{V^{N-1}}\rho_{\Sigma}(\mathbf{p})(1+\sum
\limits_{j=2}^{N}f_{1j})(1-n\int d\mathbf{r}_{2}f_{12})\label{33g}%
\end{equation}
where we approximated $\prod\limits_{j=2}^{N}(1+f_{1j})$ by $1+\sum
\limits_{j=2}^{N}f_{1j}$ for the abovementioned reasons (in view of the
further integration over the environment $N-s$ particles coordinates). In the
same way we can consider the difference between $\rho_{\Sigma}^{1}$ and
$\rho_{\Sigma}$. In the linear in $n$ approximation
\begin{equation}
\rho_{\Sigma}^{1}-\rho_{\Sigma}=\frac{1}{V^{N-1}}\rho_{\Sigma}(\mathbf{p}%
)[(\sum\limits_{j=2}^{N}f_{1j})(1-n\int d\mathbf{r}_{2}f_{12})-n\int
d\mathbf{r}_{2}f_{12}],\label{33h}%
\end{equation}
where we used that in this approximation $\rho_{\Sigma}=\frac{1}{V^{N-1}}%
\rho_{\Sigma}(\mathbf{p})$. Then, applying $\int...\int dx^{\Sigma}$ to
(\ref{33h}), we see that in the adopted approximation%
\begin{equation}
<\mathbf{F}_{1}>_{\Sigma}^{1}-<\mathbf{F}_{1}>_{\Sigma}=-n\int d\mathbf{r}%
_{2}f_{12}(\mathbf{\nabla}_{1}V_{12}).\label{33h'}%
\end{equation}

Thus, Eq. (\ref{33a}) for $s=1$ in the linear in $n$ approximation acquires
the form%
\begin{align}
\frac{\partial}{\partial t}F_{1}(t,\beta)  & =-[\mathbf{v}_{1}\mathbf{\nabla
}_{1}-(\mathbf{\nabla}_{1}<H_{1\Sigma}>_{\Sigma})\frac{\partial}%
{\partial\mathbf{p}_{1}}-n\int d\mathbf{r}_{2}f_{12}(\mathbf{\nabla}_{1}%
V_{12})\frac{\partial}{\partial\mathbf{p}_{1}}]F_{1}(t,\beta)\nonumber\\
& +C_{\operatorname{col}}(t,\beta)+C_{ic}(t,\beta),\label{33i}%
\end{align}
where $C_{\operatorname{col}}(t,\beta)$ is the collision term%

\begin{align}
C_{\operatorname{col}}(t,\beta)  & =n\int\limits_{0}^{t}d\tau\int
d\mathbf{r}_{2}\int d\mathbf{p}_{2}L_{12}^{\prime}\exp[-(L_{12}\tau
)]L_{12}^{\prime}\rho^{0}(\mathbf{p}_{2})e^{-\beta V_{12}}F_{1}(t-\tau
,\beta)\nonumber\\
& =n\int\limits_{0}^{t}d\tau\int d\mathbf{r}_{2}\int d\mathbf{p}_{2}%
\partial_{12}\mathbf{F}_{12}\exp[-(L_{12}\tau)]\mathbf{F}_{12}\partial
_{12}\rho^{0}(\mathbf{p}_{2})e^{-\beta V_{12}}\nonumber\\
& \times F_{1}(t-\tau,\beta)\nonumber\\
L_{12}  & =L_{1}^{0}+L_{2}^{0}+L_{12}^{\prime},L_{1}^{0}=\mathbf{v}%
_{1}\mathbf{\nabla}_{1},L_{2}^{0}=\mathbf{v}_{2}\mathbf{\nabla}_{2}%
,\nonumber\\
& L_{12}^{\prime}=\mathbf{F}_{12}\partial_{12},\mathbf{F}_{12}%
=-(\mathbf{\nabla}_{1}V_{12}),\partial_{12}=\frac{\partial}{\partial
\mathbf{p}_{1}}-\frac{\partial}{\partial\mathbf{p}_{2}},\nonumber\\
\rho^{0}(\mathbf{p}_{2})  & =\exp(-\beta p_{2}^{2}/2m)/\int d\mathbf{p}%
_{2}\exp(-\beta p_{2}^{2}/2m),\label{33j}%
\end{align}
and $C_{ic}(t,\beta)$ is the additional term due to initial correlations%

\begin{align}
C_{ic}(t,\beta)  & =n\int\limits_{0}^{t}d\tau\int d\mathbf{r}_{2}\int
d\mathbf{p}_{2}L_{12}^{\prime}\exp[-(L_{12}\tau)]\mathbf{g}_{12}%
(\mathbf{\nabla}_{1}f_{12})\rho^{0}(\mathbf{p}_{2})F_{1}(t-\tau,\beta
)\nonumber\\
& =n\beta\int\limits_{0}^{t}d\tau\int d\mathbf{r}_{2}\int d\mathbf{p}%
_{2}\partial_{12}\mathbf{F}_{12}\exp[-(L_{12}\tau)]\mathbf{F}_{12}%
\mathbf{g}_{12}\rho^{0}(\mathbf{p}_{2})e^{-\beta V_{12}}F_{1}(t-\tau
,\beta)\nonumber\\
\mathbf{g}_{12}  & =\mathbf{v}_{1}-v_{2}.\label{33k}%
\end{align}
Here we used Eqs. (\ref{29})-(\ref{31}), (\ref{33g}) and that in the linear in
$n$ approximation
\begin{equation}
<H_{1\Sigma}>_{\Sigma}=n\int d\mathbf{r}_{2}V_{12},\label{33l}%
\end{equation}
and that $\nabla_{1}f_{12}=-\nabla_{2}f_{12}$..

Obtained Eq. (\ref{33i}) with (\ref{33j}) and (\ref{33k}) is the main result
of this section and can be considered as a generalized linear Boltzmann
equation accounting for initial correlations. The term (\ref{33j}) is the
generalized linear version of the Boltzmann collision term. In the
space-homogeneous case when $F_{1}(t,\beta)=F_{1}(\mathbf{p}_{1};t,\beta)$,
and $\nabla_{1}<H_{1\Sigma}>_{\Sigma}=0$ and $\int d\mathbf{r}_{2}%
f_{12}(\mathbf{\nabla}_{1}V_{12})=0$, Eq. ( \ref{33i}) reduces to
\begin{equation}
\frac{\partial}{\partial t}F_{1}(\mathbf{p}_{1};t,\beta)=C_{\operatorname{col}%
}(t,\beta)+C_{ic}(t,\beta),\label{33m}%
\end{equation}
where $C_{\operatorname{col}}(t,\beta)$ and $C_{ic}(t,\beta)$ are given by
(\ref{33j}) and (\ref{33k}) but with $F_{1}(t-\tau,\beta)=F_{1}(\mathbf{p}%
_{1};t-\tau,\beta)$.

We see, that the evolution in time in Eqs. (\ref{33j}) and (\ref{33k}) is
defined by the exact two-particle propagator which satisfies the integral
equation
\begin{equation}
\exp[-(L_{12}\tau)]=U_{12}(\tau)=U_{12}^{0}(\tau)+%
%TCIMACRO{\dint \limits_{0}^{\tau}}%
%BeginExpansion
{\displaystyle\int\limits_{0}^{\tau}}
%EndExpansion
d\tau_{1}U_{12}^{0}(\tau-\tau_{1})L_{12}^{^{\prime}}U_{12}(\tau_{1}%
),\label{33n}%
\end{equation}
where $U_{12}^{0}(\tau)=\exp[-(L_{1}^{0}+L_{2}^{0})\tau]$ is a "free"
two-particle propagator.

\section{Connection to the Boltzmann equation}

Note, that Eqs. (\ref{33i}) and (\ref{33m}) are the reversible in time ones.
They become irreversible if it is possible to extend in them the upper limit
of integration over $\tau$ to infinity (and this limit exists). It can be the
case of the short time of interparticle interaction $\tau_{cor}$, when%
\begin{equation}
\tau_{cor}<<t\thicksim\tau_{rel},\label{33o}%
\end{equation}
where $\tau_{rel}$ is the timescale on which the $F_{1}(t,\beta)$ changes (see
also below). Then, in the obtained equations we can approximate the
one-particle function as $F_{1}(t-\tau,\beta)\thickapprox F_{1}(t,\beta)$\ and
simultaneously extend the upper limit of integration to infinity. The latter
can be done, e.g., if the time-dependent force-force correlation function in
Eqs. (\ref{33j}) and (\ref{33k})%
\begin{equation}
\int d\mathbf{r}_{2}\mathbf{F}_{12}\exp[-(L_{12}\tau)]\mathbf{F}%
_{12}\label{33p}%
\end{equation}
$\pi$ quickly vanishes on the timescale $\tau_{cor}\sim r_{cor}/v<<\tau_{rel}$
($r_{cor}$ is a radius of the inter-particle interaction $V_{ij}$ and $v$ is
the average particle velocity), i.e. when the interaction is rather a
short-range one. Thus, in this case, Eqs. (\ref{33i}) and (\ref{33m}) on the
timesacle $t\thicksim\tau_{rel}$ become Markovian (time local) and irreversible.

Let us consider (following the approach of \cite{Balescu}) in this
approximation and in the space-homogeneous case the collision term
(\ref{33j}), which in new convenient variables $\mathbf{v}_{i}=\mathbf{p}%
_{i}/m$, $\mathbf{r}=\mathbf{r}_{2}-\mathbf{r}_{1}$ and $\mathbf{g}%
=\mathbf{v}_{1}-\mathbf{v}_{2}$ can be written as%
\begin{align}
C_{\operatorname{col}}(\mathbf{v}_{1};t,\beta)  & =n\int d\mathbf{v}%
_{2}J(\mathbf{v}_{1},\mathbf{v}_{2}),\nonumber\\
J(\mathbf{v}_{1},\mathbf{v}_{2})  & =\int d\mathbf{r}\int\limits_{0}^{\infty
}d\tau L^{\prime}U(\tau)L^{\prime}\varphi(\mathbf{v}_{1},\mathbf{v}%
_{2},\mathbf{r};t,\beta),\nonumber\\
L^{\prime}  & =[\mathbf{\nabla}V(\mathbf{r})]\ast\mathbf{\partial,\nabla
=}\frac{\partial}{\partial\mathbf{r}},\partial=\frac{2}{m}\frac{\partial
}{\partial\mathbf{g}},\nonumber\\
\varphi(\mathbf{v}_{1},\mathbf{v}_{2},\mathbf{r};t,\beta)  & =\rho
^{0}(\mathbf{v}_{2})e^{-\beta V(\mathbf{r})}F_{1}(\mathbf{v}_{1}%
;t,\beta),\label{33q}%
\end{align}
where we dropped for brevity the indexes $12$ in (\ref{33j}), which cannot
lead to misunderstanding because we deal in the adopted first order in $n$
approximation only with a pair of particles. It can be shown \cite{Balescu},
that the integral over $\tau$ in (\ref{33q}) can be presented as (see
(\ref{33n}))
\begin{equation}
Z=G+GL^{\prime}Z,G=%
%TCIMACRO{\dint \limits_{0}^{\infty}}%
%BeginExpansion
{\displaystyle\int\limits_{0}^{\infty}}
%EndExpansion
d\tau U^{0}(\tau),Z=%
%TCIMACRO{\dint \limits_{0}^{\infty}}%
%BeginExpansion
{\displaystyle\int\limits_{0}^{\infty}}
%EndExpansion
d\tau U(\tau).\label{33s}%
\end{equation}
In the matrix form, $Z(\mathbf{r},\mathbf{g};\mathbf{r}^{\prime}%
,\mathbf{g}^{\prime})$ satisfies the equation
\begin{equation}
\{\mathbf{g\nabla-}[\mathbf{\nabla}V(\mathbf{r})]\ast\mathbf{\partial
}\}Z(\mathbf{r},\mathbf{g};\mathbf{r}^{\prime},\mathbf{g}^{\prime}%
)=\delta(\mathbf{r}-\mathbf{r}^{\prime})\delta(\mathbf{g}-\mathbf{g}^{\prime
}),\label{33t}%
\end{equation}
i.e., it is the Green function of the two-particle Liouville equation (see
(\ref{31})), whereas the matrix $G(\mathbf{r},\mathbf{g};\mathbf{r}^{\prime
},\mathbf{g}^{\prime})$ is diagonal with respect to velocity indexes
$G(\mathbf{r},\mathbf{g};\mathbf{r}^{\prime},\mathbf{g}^{\prime}%
)=G^{0}(\mathbf{r}-\mathbf{r}^{\prime})\delta(\mathbf{g}-\mathbf{g}^{\prime})$
and $G^{0}(\mathbf{r},\mathbf{g};\mathbf{r}^{\prime},\mathbf{g}^{\prime})$ is
the Green function of the unperturbed Liouville equation, i.e.,%
\begin{equation}
\mathbf{g\nabla}G^{0}(\mathbf{r}-\mathbf{r}^{\prime})=\delta(\mathbf{r}%
-\mathbf{r}^{\prime}).\label{33u}%
\end{equation}
If we introduce the function%
\begin{equation}
f(\mathbf{r},\mathbf{g;}t,\beta)=\varphi(\mathbf{v}_{1},\mathbf{v}%
_{2},\mathbf{r};t,\beta)+\int d\mathbf{r}^{\prime}\int d\mathbf{g}^{\prime
}Z(\mathbf{r},\mathbf{g};\mathbf{r}^{\prime},\mathbf{g}^{\prime}%
)[\mathbf{\nabla}^{\prime}V(\mathbf{r}^{\prime})]\mathbf{\partial}^{\prime
}\varphi(\mathbf{v}_{1}^{\prime},\mathbf{v}_{2}^{\prime},\mathbf{r}^{\prime
};t,\beta),\label{33v}%
\end{equation}
then it is not difficult to show, using (\ref{33t}), that%
\begin{align}
\{\mathbf{g\nabla-}[\mathbf{\nabla}V(\mathbf{r})]\ast\mathbf{\partial
\}}f(\mathbf{r},\mathbf{g;}t,\beta)  & =\mathbf{g\nabla}\varphi(\mathbf{v}%
_{1},\mathbf{v}_{2},\mathbf{r};t,\beta),\nonumber\\
\lim_{V(r)\rightarrow0}f(\mathbf{r},\mathbf{g;}t,\beta)  & =\rho
^{0}(\mathbf{v}_{2})F_{1}(\mathbf{v}_{1};t,\beta).\label{33w}%
\end{align}
Thus, it is easy to verify that function $f(\mathbf{r},\mathbf{g;}t,\beta)$,
presented as%
\begin{equation}
f(\mathbf{r},\mathbf{g;}t,\beta)=\varphi(\mathbf{v}_{1},\mathbf{v}%
_{2},\mathbf{r};t,\beta)+\int d\mathbf{r}^{\prime}G^{0}(\mathbf{r}%
-\mathbf{r}^{\prime})[\mathbf{\nabla}^{\prime}V(\mathbf{r}^{\prime
})]\mathbf{\partial}f(\mathbf{r}^{\prime},\mathbf{g;}t,\beta),\label{33x}%
\end{equation}
satisfies the equation (\ref{33w}).

Now we can write down the function $J(\mathbf{v}_{1},\mathbf{v}_{2})$ defining
the collision term (\ref{33q}) as%
\begin{align}
J(\mathbf{v}_{1},\mathbf{v}_{2})  & =\int d\mathbf{r}[\mathbf{\nabla
}V(\mathbf{r})]\mathbf{\partial\lbrack}f\mathbf{(\mathbf{r},\mathbf{g;}%
}t,\beta\mathbf{)-}\varphi\mathbf{(\mathbf{v}_{1},\mathbf{v}_{2},\mathbf{r}%
;}t,\beta\mathbf{)]}\nonumber\\
& =\int d\mathbf{rg\nabla}V(\mathbf{r})f\mathbf{(\mathbf{r},\mathbf{g;}%
}t,\beta\mathbf{).}\label{33y}%
\end{align}
Here we used (\ref{33v}), (\ref{33w}) and that $\varphi\mathbf{(\mathbf{v}%
_{1},\mathbf{v}_{2},\mathbf{r};}t,\beta\mathbf{)}$ depends on the relative
distance $r$ as $\exp[-\beta V(r)]$, and, therefore, $\int d\mathbf{r}%
[\mathbf{\nabla}V(\mathbf{r})]\exp[-\beta V(r)]=-\beta^{-1}\int
d\mathbf{r\nabla}\{\exp[-\beta V(r)]\}=0$ in the space-homogeneous case.

Let us select the coordinate system in which axis $z$ is directed along vector
$\mathbf{g}$. Then the Green function $G^{0}(\mathbf{r}-\mathbf{r}^{\prime})$
has the form
\begin{align}
G^{0}(\mathbf{r}-\mathbf{r}^{\prime})  & =g^{-1}\delta(x-x^{\prime}%
)\delta(y-y^{\prime})\theta(z-z^{\prime}),\nonumber\\
\theta(x)  & =1,x>0,\nonumber\\
\theta(x)  & =0,x<0,\label{33z}%
\end{align}
which is in agreement with Eq. (\ref{33u}). Inserting (\ref{33z}) in
(\ref{33x}), we obtain%
\begin{align}
f(\mathbf{r},\mathbf{g;}t,\beta)  & =\varphi(\mathbf{v}_{1},\mathbf{v}%
_{2},\mathbf{r};t,\beta)+%
%TCIMACRO{\dint \limits_{-\infty}^{z}}%
%BeginExpansion
{\displaystyle\int\limits_{-\infty}^{z}}
%EndExpansion
dz^{\prime}g^{-1}[\nabla^{\prime}V(x,y,z^{\prime})]\mathbf{\partial
}f(x,y,z^{\prime},g\mathbf{;}t,\beta)\nonumber\\
& =\varphi(\mathbf{v}_{1},\mathbf{v}_{2},\mathbf{r};t,\beta)+%
%TCIMACRO{\dint \limits_{-\infty}^{z}}%
%BeginExpansion
{\displaystyle\int\limits_{-\infty}^{z}}
%EndExpansion
dz^{\prime}(\frac{\partial}{\partial z^{\prime}})[f(x,y,z^{\prime}%
,g\mathbf{;}t,\beta)-\varphi(\mathbf{v}_{1},\mathbf{v}_{2},z^{\prime}%
;t,\beta)]\nonumber\\
& =\varphi(\mathbf{v}_{1},\mathbf{v}_{2},\mathbf{r};t,\beta)+%
%TCIMACRO{\dint \limits_{-\infty}^{z}}%
%BeginExpansion
{\displaystyle\int\limits_{-\infty}^{z}}
%EndExpansion
dz^{\prime}(\frac{\partial}{\partial z^{\prime}})f(x,y,z^{\prime}%
,g\mathbf{;}t,\beta)\nonumber\\
& =\varphi(\mathbf{v}_{1},\mathbf{v}_{2},\mathbf{r};t,\beta)+\Phi
(x,y,z,g;t,\beta),\nonumber\\
\Phi(x,y,z,g\mathbf{,}\beta)  & =%
%TCIMACRO{\dint \limits_{-\infty}^{z}}%
%BeginExpansion
{\displaystyle\int\limits_{-\infty}^{z}}
%EndExpansion
dz^{\prime}(\frac{\partial}{\partial z^{\prime}})f(x,y,z^{\prime}%
,g\mathbf{;}t,\beta).\label{33A}%
\end{align}
Finally, introducing (\ref{33A}) in (\ref{33y}), we obtain%
\begin{align}
J(\mathbf{v}_{1},\mathbf{v}_{2})  & =\int d\mathbf{rg\nabla}V(\mathbf{r}%
)f\mathbf{(\mathbf{r},\mathbf{g;}}t,\beta\mathbf{)}\nonumber\\
& =%
%TCIMACRO{\dint \limits_{-\infty}^{\infty}}%
%BeginExpansion
{\displaystyle\int\limits_{-\infty}^{\infty}}
%EndExpansion
dzg(\frac{\partial}{\partial z})\Phi(x,y,z)=g[\Phi(x,y,\infty,g;t,\beta
)-\Phi(x,y,-\infty,g;t,\beta)]\nonumber\\
& =g%
%TCIMACRO{\dint \limits_{-\infty}^{\infty}}%
%BeginExpansion
{\displaystyle\int\limits_{-\infty}^{\infty}}
%EndExpansion
dz^{\prime}(\frac{\partial}{\partial z^{\prime}})f(x,y,z^{\prime}%
,g\mathbf{;}t,\beta)\nonumber\\
& =g[f(x,y,+\infty,g\mathbf{;}t,\beta)-f(x,y,-\infty,g\mathbf{;}%
t,\beta)].\label{33B}%
\end{align}

It is then evident from (\ref{33B}), that function $f(x,y,-\infty
,g\mathbf{;}t,\beta)$ can be identified with the distribution function
(\ref{33q}) $\varphi(\mathbf{v}_{1},\mathbf{v}_{2},\mathbf{-\infty};t,\beta)$
prior to collision, i.e.%
\begin{equation}
\varphi(\mathbf{v}_{1},\mathbf{v}_{2},\mathbf{-\infty};t,\beta)=\rho
^{0}(\mathbf{v}_{2})F_{1}(\mathbf{v}_{1};t,\beta),\label{33C}%
\end{equation}
where we used that $\exp[-\beta V(\pm\infty)]=1$ ($V(\pm\infty)=0$). Function
$f(x,y,+\infty,g\mathbf{;}t,\beta)$ can be considered as the distribution
function of particles after collision with the relative velocity $\mathbf{g}$.
But according to the Liouville theorem, this distribution function is equal to
the distribution function before collision with velocities $\mathbf{v}%
_{1}^{\prime}$, $\mathbf{v}_{2}^{\prime}$ which correspond to the velocities
$\mathbf{v}_{1}$, $\mathbf{v}_{2}$, i.e.,
\begin{align}
f(x,y,+\infty,g\mathbf{;}t,\beta)  & =\rho^{0}(\mathbf{v}_{2}^{\prime}%
)F_{1}(\mathbf{v}_{1}^{\prime};t,\beta),\nonumber\\
\mathbf{v}_{1}+\mathbf{v}_{2}  & =\mathbf{v}_{1}^{\prime}+\mathbf{v}%
_{2}^{\prime},\nonumber\\
v_{1}^{2}+v_{2}^{2}  & =v_{1}^{\prime2}+v_{2}^{\prime2}.\label{33D}%
\end{align}
Taking into account that in the adopted coordinate system with $\mathbf{g}$
directed along $z$-axis%
\begin{equation}
dxdy=bdbd\varphi,\label{33D'}%
\end{equation}
where $b$ is the impact parameter and $\varphi$ is the azimuth angle. As a
result, we arrive at the linear Boltzmann collision term in Eq. (\ref{33m})%
\begin{equation}
C_{\operatorname{col}}(\mathbf{v}_{1};t,\beta)=n\int d\mathbf{v}_{2}\int
d\varphi dbbg[\rho^{0}(\mathbf{v}_{2}^{\prime})F_{1}(\mathbf{v}_{1}^{\prime
};t,\beta)-\rho^{0}(\mathbf{v}_{2})F_{1}(\mathbf{v}_{1};t,\beta).\label{33E}%
\end{equation}
Making in (\ref{33E}) the substitutions
\begin{equation}
F_{1}(\mathbf{v}_{1};t,\beta)=\rho^{0}(\mathbf{v}_{1})W(\mathbf{v}%
_{1},t),F_{1}(\mathbf{v}_{1}^{\prime};t,\beta)=\rho^{0}(\mathbf{v}_{1}%
^{\prime})W(\mathbf{v}_{1}^{\prime},t),\label{33F}%
\end{equation}
the collision term for the function $W$ can be rewritten as (see, e.g.
\cite{Catapano})%
\begin{equation}
C_{\operatorname{col}}(\mathbf{v}_{1};t,\beta)=\rho^{0}(\mathbf{v}_{1})n\int
d\mathbf{v}_{2}\int d\varphi dbbg\rho^{0}(\mathbf{v}_{2})[W(\mathbf{v}%
_{1}^{\prime},t)-W(\mathbf{v}_{1},t)],\label{33G}%
\end{equation}
where we used that according to definition (\ref{33j}) for $\rho
^{0}(\mathbf{v}_{2})$ and the conservation of energy (\ref{33D})%
\begin{equation}
\rho^{0}(\mathbf{v}_{1})\rho^{0}(\mathbf{v}_{2})=\rho^{0}(\mathbf{v}%
_{1}^{\prime})\rho^{0}(\mathbf{v}_{2}^{\prime})\label{33H}%
\end{equation}
Note, that $\rho^{0}(\mathbf{v}_{1})$ can be cancelled out in Eq. (\ref{33m})
written for function $W(\mathbf{v}_{1},t)$.

\section{Connection to the Landau equation}

It is interesting to consider Eq. (\ref{33i}) for the case of a weak
interparticle interaction when
\begin{equation}
<p_{i}^{2}/2m>\backsim k_{B}T>>V_{ij},\label{34}%
\end{equation}
i.e., the interparticle interaction is small as compared to the average
particle's kinetic energy. Then, in the second order in the small parameter,
defined by inequality (\ref{34}), the collision term (\ref{33j}) can be
rewritten as%
\begin{equation}
C_{\operatorname{col}}(t,\beta)=n\int\limits_{0}^{t}d\tau\int d\mathbf{r}%
_{2}\int d\mathbf{p}_{2}\mathbf{F}_{12}\mathbf{\partial}_{12}\exp
[-(\mathbf{v}_{1}\mathbf{\nabla}_{1}+\mathbf{v}_{2}\mathbf{\nabla}_{2}%
)\tau]\mathbf{F}_{12}\partial_{12}\rho^{0}(\mathbf{p}_{2})e^{\mathbf{v}%
_{1}\mathbf{\nabla}_{1}\tau}F_{1}(t,\beta).\label{34a}%
\end{equation}
Here, in order to remain within adopted accuracy, we took $F_{1}(t-\tau
,\beta)$ in the zero in the interaction approximation (see (\ref{33i}))%
\begin{equation}
F_{1}(t-\tau,\beta)=\exp[-\mathbf{v}_{1}\mathbf{\nabla}_{1}(t-\tau
)]F_{1}(0,\beta)=e^{\mathbf{v}_{1}\mathbf{\nabla}_{1}\tau}F_{1}(t,\beta
)\label{34b}%
\end{equation}
In the same approximation%
\begin{equation}
C_{ic}(t,\beta)r\mathbf{r}_{2}\int d\mathbf{p}_{2}\mathbf{F}_{12}\partial
_{12}\exp[-(\mathbf{v}_{1}\mathbf{\nabla}_{1}+\mathbf{v}_{2}\mathbf{\nabla
}_{2})\tau]\mathbf{F}_{12}\mathbf{g}_{12}\rho^{0}(\mathbf{p}_{2}%
)e^{\mathbf{v}_{1}\mathbf{\nabla}_{1}\tau}F_{1}(t,\beta).\label{34c}%
\end{equation}

Then, we have for any function of the particles coordinates $\Phi
(\mathbf{r}_{1},\mathbf{r}_{2},...,\mathbf{r}_{N})$
\begin{align}
&  \exp[-(\mathbf{v}_{1}\mathbf{\nabla}_{1}+\mathbf{v}_{2}\mathbf{\nabla}%
_{2})\tau\mathbf{]}\Phi(\mathbf{r}_{1},\mathbf{r}_{2},..,\mathbf{r}%
_{N})\nonumber\\
&  =\Phi(\mathbf{r}_{1}-\mathbf{v}_{1}\tau,\mathbf{r}_{2}-\mathbf{v}_{2}%
\tau,...,\mathbf{r}_{N}).\label{34d}%
\end{align}
If we also take into account the commutation rule%
\begin{equation}
\left[  \exp[-(\mathbf{v}_{1}\mathbf{\nabla}_{1}+\mathbf{v}_{2}\mathbf{\nabla
}_{2})\tau],\partial_{12}\right]  =\exp[-(\mathbf{v}_{1}\mathbf{\nabla}%
_{1}+\mathbf{v}_{2}\mathbf{\nabla}_{2})\tau]\frac{\tau}{m}(\mathbf{\nabla}%
_{1}-\mathbf{\nabla}_{2}),\label{34e}%
\end{equation}
the collision term acquires the final form%

\begin{align}
C_{\operatorname{col}}(t,\beta)  & =n\int\limits_{0}^{t}d\tau%
%TCIMACRO{\dint }%
%BeginExpansion
{\displaystyle\int}
%EndExpansion
d\mathbf{p}_{2}\mathbf{\partial}_{12}G_{L}(x_{1},\mathbf{g}_{12}\mathbf{;}%
\tau)(\partial_{12}+\frac{\tau}{m}\mathbf{\nabla}_{1})\rho^{0}(\mathbf{p}%
_{2})F_{1}(x_{1};t,\beta),\nonumber\\
G_{L}(\mathbf{r}_{1},\mathbf{g}_{12}\mathbf{;}\tau)  & =%
%TCIMACRO{\dint }%
%BeginExpansion
{\displaystyle\int}
%EndExpansion
d\mathbf{r}_{2}\mathbf{F}_{12}(0)\mathbf{F}_{12}(\tau),\mathbf{F}_{12}%
(\tau)=-\mathbf{\nabla}_{1}V(\mathbf{r}_{1}-\mathbf{r}_{2}-\mathbf{g}_{12}%
\tau).\label{34f}%
\end{align}
Using (\ref{34d}), the initial correlation term (\ref{34c}) can be rewritten
as
\begin{equation}
C_{ic}(t,\beta)=n\beta\int\limits_{0}^{t}d\tau\int d\mathbf{p}_{2}%
\partial_{12}G_{L}(x_{1},\mathbf{g}_{12}\mathbf{;}\tau)\mathbf{g}_{12}\rho
^{0}(\mathbf{p}_{2})F_{1}(x_{1};t,\beta).\label{34g}%
\end{equation}

The collision integral (\ref{34f}) coincides with the corresponding collision
integral in the nonlinear equation for inhomogeneous system of weakly
interacting classical particles (see \cite{Balescu}) if in the latter, the
distribution function for the particle, with which the tagged particle
collides, is replaced by the equilibrium distribution function for this
particle $\rho^{0}(\mathbf{p}_{2})$. In addition, for such a coincidence, the
integral over $d\tau$ should be extended to infinity. It can be done, if the
interaction is rather a short-range one and if for the timescale (\ref{33o})
the force acting on the particle vanishes ($\mathbf{F}_{12}(t)=0$).

In the space-homogeneous case equation (\ref{33m}) for a small interparticle
interaction reads%
\begin{equation}
\frac{\partial}{\partial t}F_{1}(\mathbf{p}_{1};t,\beta)=n\int\limits_{0}%
^{t}d\tau%
%TCIMACRO{\dint }%
%BeginExpansion
{\displaystyle\int}
%EndExpansion
d\mathbf{p}_{2}\mathbf{\partial}_{12}G_{L}(x_{1},\mathbf{g}_{12}\mathbf{;}%
\tau)(\partial_{12}+\beta\mathbf{g}_{12})\rho^{0}(\mathbf{p}_{2}%
)F_{1}(\mathbf{p}_{1};t,\beta).\label{34h}%
\end{equation}
The first (collision) term in the r.h.s. of Eq. (\ref{34h}) coincides with the
Landau collision integral if it is possible to extend integral over $\tau$ to
infinity (short-range interparticle interaction) and to replace the
distribution function for the second tagged particle with $\rho^{0}%
(\mathbf{p}_{2})$ (which seems natural for considered second order in
interaction approximation for the kernel governing evolution of $F_{1}%
(\mathbf{p}_{1};t,\beta)$). The second term in the r.h.s. of (\ref{34h}) is
caused by initial correlations.

\section{Conclusion}

We have rigorously derived the exact completely closed (homogeneous)
Generalized Master Equations governing the evolution in time of an equilibrium
two-time correlation function for dynamic variables of a selected group of $s$
($s<N$) particles of an $N$-particle ($N>>1$) system of classical particles.
These time-convolution and time-convolutionless GMEs differ from known
equations (such as Nakajima-Zwanzig equation) by absence of the undesirable
inhomogeneous terms containing the correlations of all $N$ particles in the
initial moment of time. Such reduced description has become possible due to
employing a special projection operator.

This projection operator (comprising initial correlations) can be expanded
into series in the density of particles or in the weak interparticle
interaction. In the linear in $n$ approximation for the kernel governing the
evolution of a one-particle correlation function, the generalized linear
Boltzmann equation accounting for initial correlations and valid at any
timescale has been rigorously obtained. At the timescale $t\thicksim\tau
_{rel}>>\tau_{cor}$ and for a short-range interaction this equation becomes
irreversible with the collision integral of the known linear Boltzmann
equation but with additional term due to initial correlations.

If in addition the interparticle interaction is weak, the generalized linear
Boltzmann equation converts into the generalized linear Landau equation
accounting for initial correlations and valid on all timescale. Again, at the
timescale $t\thicksim\tau_{rel}>>\tau_{cor}$ and for a short-range interaction
this equation becomes irreversible. For the space-homogeneous case, the
collision integral coincides with the Landau collision integral in which the
distribution function of the second tagged particle is replaced with the
equilibrium Maxwell distribution function. But there is also an additional
term in the kernel governing the evolution of a one-particle correlation
function caused by initial correlations.

\bigskip

\begin{thebibliography}{99}                                                                                               %
\bibitem {Bogoliubov}N. N. Bogoliubov, \textit{Problems of Dynamical Theory in
Statistical Physics\ }(Gostekhizdat, Moscow, 1946, in Russian); English
transl.: \textit{Stud. Statist. Mech.\ }\textbf{1} (North-Holland, Amsterdam, 1962).

\bibitem {Lanford (1975)}O. E. Lanford, \textit{Time evolution of large
classical systems.} \textit{In "Dynamical systems, theory and applications",}
Lecture Notes in Physics, ed. J. Moser, \textbf{38}, 1-111 (Springer-Verlag,
Berlin, 1975).

\bibitem {Balescu}R. Balescu,\textit{\ Equilibrium and Nonequilibrium
Statistical Mechanics\ }(Wiley-Interscience, New York, 1975).

\bibitem {Bobylev (2013)}A. Bobylev, M. Pulvirenti and C. Saffirio, Comm.
Math. Phys. \textbf{319}, 683 (2013).

\bibitem {Breuer}H. -P. Breuer and F. Petruccione, \textit{The Theory of Open
Quantum Systems} (Oxford University Press, New York, 2002).

\bibitem {Van Kampen}N. G. van Kampen, J. Stat. Phys. \textbf{115, }1057 (2004).

\bibitem {Kac}M. Kac, \textit{Probability and related topics in physical
sciences} (Interscience, London-New York, 1959).

\bibitem {Los JPA}V. F. Los, J. Phys. A: Math. Gen. \textbf{34, }6389 (2001).

\bibitem {Los JSP}V. F. Los, J. Stat. Phys. \textbf{119, }241 (2005).

\bibitem {Los Evolution equations}V. F. Los, in \textit{Evolution Equations,
}edited by A. Claes (Nova Science Publishers, New York, 2012), p. 251.

\bibitem {Los Theor. Math. Phys}V. F. Los, Theor. Math. Phys. \textbf{160,
}1124 (2009).

\bibitem {Los 2017}V. F. Los, J. Stat. Phys. \textbf{168}, 857
(2017\textbf{).}

\bibitem {Los 2018}V. F. Los, Physica A \textbf{503}, 476 (2018).

\bibitem {Nakajima (1958)}S. Nakajima, Progr. Theor. Phys. \textbf{20}, 948 (1958).

\bibitem {Zwanzig (1960)}R. Zwanzig, J. Chem. Phys. \textbf{33}, 1338 (1960).

\bibitem {Prigogine (1962)}I. Prigogine, \textit{Non-Equilibrium Statistical
Mechanics} (Interscience Publishers, New York, 1962).

\bibitem {Shibata1 (1977)}F. Shibata, Y. Takahashi, and N. Hashitsume, J.
Stat. Phys. \textbf{17}, 171 (1977).

\bibitem {Shibata2 (1980)}F. Shibata and T. Arimitsu, J. Phys. Soc. Jap.
\textbf{49}, 891 (1980).

\bibitem {Mayer}J. E. Mayer and E. W. Montroll, J. Chem. Phys. \textbf{9}, 2 (1941).

\bibitem {Catapano}N. Catapano, Kinetic and Related Models, \textbf{11(3),}
647 (2018), arXiv: 1701.05465.
\end{thebibliography}
\end{document}